# Nanoscale three-dimensional magnetic sensing with a probabilistic nanomagnet driven by spin-orbit torque

Shuai Zhang[1]†, Shihao Li[1]†, Zhe Guo[1], Yan Xu[1], Ruofan Li[1], Zhenjiang Chen[1], Min Song[2], Xiaofei Yang[1], Liang Li[3], Jeongmin Hong[1], Xuecheng Zou[1], Long You[1,3]*

**Affiliations:**

[1]School of Optical and Electronic Information & Wuhan National Laboratory for Optoelectronics, Huazhong University of Science and Technology; Wuhan, 430074, China

[2]Faculty of Physics and Electronic Science, Hubei University; Wuhan, 430062, China

[3]Wuhan National High Magnetic Field Center, Huazhong University of Science and Technology; Wuhan, 430074, China

*Corresponding author. Email: lyou@hust.edu.cn

†These authors contributed equally to this work.

**Abstract:** Detection of vector magnetic fields at nanoscale dimensions is critical in applications ranging from basic material science, to medical diagnostic. Meanwhile, an all-electric operation is of great significance for achieving a simple and compact sensing system. Here, we propose and experimentally demonstrate a simple approach to sensing a vector magnetic field at nanoscale dimensions, by monitoring a probabilistic nanomagnet's transition probability from a metastable state, excited by a driving current due to SOT, to a settled state. We achieve sensitivities for $H_x$, $H_y$, and $H_z$ of 1.02%/Oe, 1.09%/Oe and 3.43%/Oe, respectively, with a 200×200 nm$^2$ nanomagnet. The minimum detectable field is dependent on the driving pulse events $N$, and is expected to be as low as 1 μT if $N = 3 \times 10^6$.

**Main Text:** Sensitive physical measurements have played a pivotal role in the development of modern science and technology. In particular, detection of vectorial magnetic fields with high sensitivity and nanoscale spatial resolution is critical important in fundamental studies ranging from physics and material sciences to biology, as well as in the development of new applications in spintronics and quantum technology (*1, 2*). Consequently, over the past few decades, various advanced magnetic field sensors have been developed, including scanning Hall probe microscopy (*3*), nanoscale superconducting interference devices (*4*, *5*), magnetic resonance force microscopy (*6, 7*), giant magnetoresistance, tunnelling magnetoresistance (TMR) (*8-10*), and nitrogen vacancy (NV) center in single-crystal diamond (*11-13*) to achieve nanometer resolution. Unfortunately, such technologies can solely measure the magnitude of the component of magnetic field that lies along its sensitive axis. One conventional method for measuring a vector field is to use three magnetic sensors (or NV ensembles) with their sensing directions along the three coordinate axes (*x*, *y*, and *z*), which degrades the spatial resolution (*14*-*16*).

Recently, single NV have been proposed as a magnetic vectorial detecting system to provide the nanoscale spatial resolution, but such a system requires precisely fabricating a single crystal diamond with an NV center as well as additional lasers and microwave excitation, which limit the practical applications and scalability of the approach (*17, 18*). On the other hand, a single spintronic device can implement the detection of three-dimensional (3D) magnetic field by the spin-orbit torque (SOT), which offers advantages such as excellent magnetic sensitivity, fast measurement, complementary metal-oxide-semiconductor (CMOS) process compatibility, and all-electric operating environment at room temperature (*19*, *20*). However, the device size cannot be unrestricted scaled down to maintain the multidomain structure. Therefore, there is a need for a single spin-dependent device that can detect a vectoral magnetic field with extremely high spatial resolution.

Meanwhile, if a nanomagnet, composed of heavy metal (HM)/ferromagnet (FM) heterostructure such as Ta/CoFeB/MgO, has perpendicular magnetic anisotropy (PMA), SOT induced by an in-plane (IP) current flowing in the HM (Ta) layer can drive the magnetization of nanomagnet to probabilistically switch, and the switching probability is dependent on the applied IP magnetic fields (*21, 22*). These results indicate it may be feasible to detect a 2D and even 3D magnetic field, through monitoring the switching probability of the nanomagnet. To be specific, under a current pulse, the magnetization of the nanomagnet (such as CoFeB) can be excited from a stable state ('up' or 'down' state, named $M_z = \pm 1$) to a metastable state (in-plane state, named $M_z = 0$) via SOT (*22*-*24*). When the magnetization relaxes to a stable state once the current pulse is turned off, the transition probability from $M_z = 0$ to $M_z = \pm 1$ differs depending on the Zeeman splitting of $M_z = \pm 1$, under an external or effective magnetic field along the out-of-plane ($z$) direction (*25*-*27*). The IP ($x$ or $y$) components of the magnetic field together with the collinear current can generate and modulate the $z$-direction effective field due to SOT, and the sign of this field depends on the current polarity (*24*). This phenomenon enables the IP and OOP components to be distinguished with alternating current pulse directions ($\pm x$ or $\pm y$) (*19*). In this work, we demonstrate that such a compact and simple approach is indeed possible for nanoscale vector magnetic field detecting, with a probabilistic nanomagnet composed of HM/FM heterostructure, driven by SOT.

## Sensing principle and experimental setup

A single domain nanomagnet with PMA sitting immediately on a Ta layer is the basic field probing element of the proposed sensor, as illustrated in Fig.1A. In this configuration, the anomalous Hall resistance $R_{AHE}$ is proportional to the out-of-plane magnetization, according to the anomalous Hall effect (AHE), and thus one can probe the magnetization state by measuring the $R_{AHE}$. As a result of the spin Hall effect of Ta or Rashba effect in the Ta/CoFeB/MgO heterostructure, the magnetization of nanomagnet can be controlled by charge current flowing in the HM layer via SOT (*28*, *29*). The in-plane torque, if strong enough, can bring nanomagnet's magnetization to the metastable state ($M_z = 0$) at an energy level of $E_b$ (energy barrier) (*22*-*24*). Additionally, after relax, from a macrospin perspective, the magnetization state will be found in the lower energy state with higher probability, where the probability of being in a state with energy $E$ is proportional to $\exp(-E/k_BT)$ with $k_B$ the Boltzmann constant and $T$ the temperature, as given by Boltzmann statistics (*30*, *31*) or deduced by the Arrhenius-Néel law (*25*) (see Supplementary Section 1 for more details). Note that the degenerated energy level of stable state ($M_z = \pm 1$) is assumed to be ground state ($E = 0$) without any effective magnetic field (Fig. 1B).

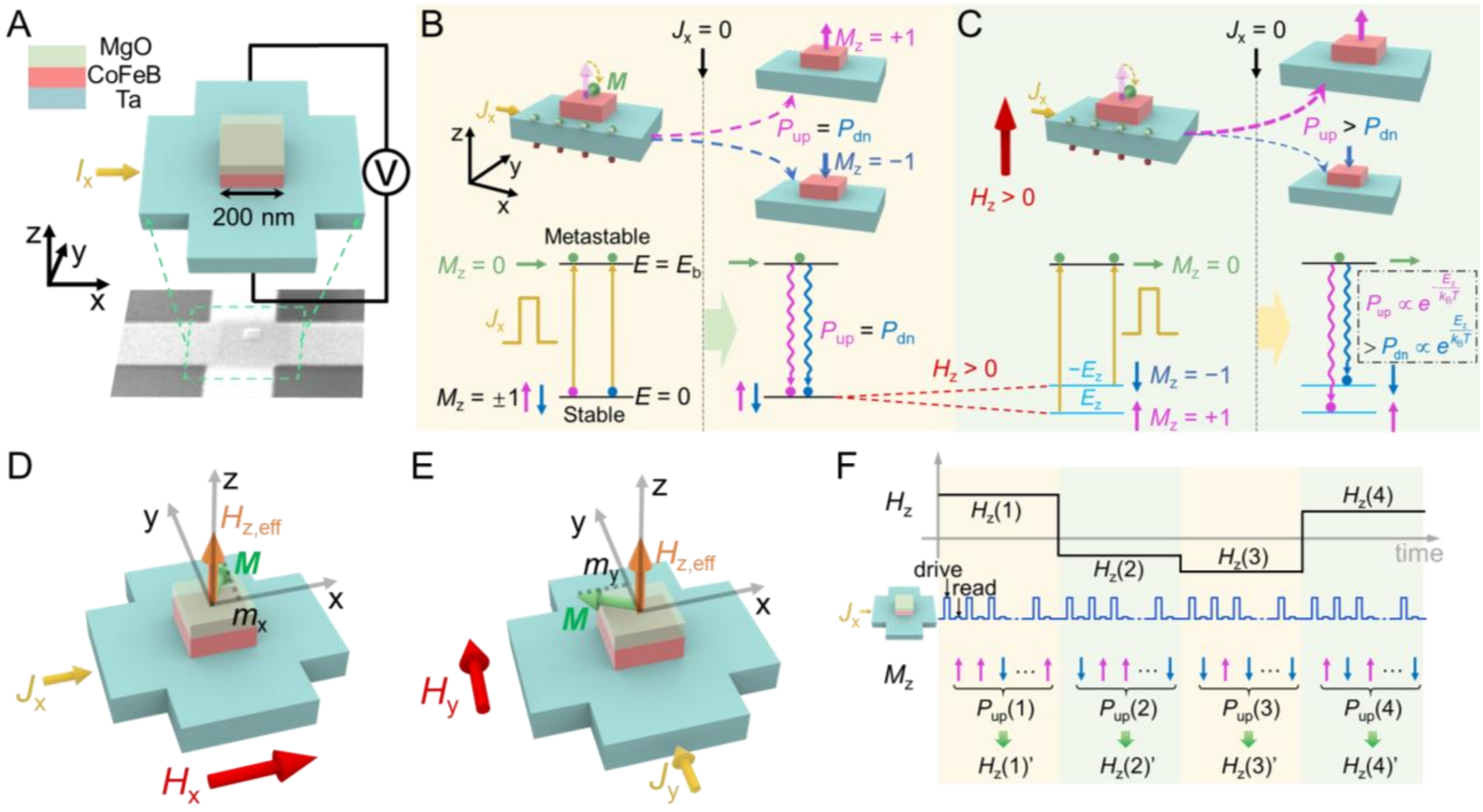

**Fig. 1. Schematic visualization of the bilayer nanomagnet sensing principle.** (**A**) Device structure along with the schematic of the measurement setup (top) and the SEM image of the nanomagnet (bottom). The current is applied along the *x*-direction while the Hall voltage is detected in the *y*-direction. (**B**) Energy level structure of the bistate nanomagnet with PMA under zero field. The energies for being in the upward and downward magnetized states ($M_z = \pm1$) are the same ($E = 0$). If the magnet is stimulated by a driving current, then the magnetization will be excited from the stable state to the metastable state ($M_z = 0$) with an energy level of $E_b$. Once the current is turned off ($J_x = 0$), the magnetization will return to the up or down state with equal probability. (**C**) Energy level structure of the nanomagnet under a field of $+H_z$. The energies of the up and down states are split by $E_z$ and $-E_z$, respectively. After the driving current is turned off, the magnetization will be more likely to relax to the state with lower energy, resulting in $P_{up} > P_{dn}$. (**D** and **E**) $H_{z,eff}$ produced by SOT when the current and the colinear field are along *x*-direction (D) and *y*-direction (E). (**F**) Diagram of the approach for sensing an OOP field. Under a certain OOP magnetic field, a batch of current pulses are applied to the device along the *x*-direction to count the state numbers and thus calculate the corresponding probability ($P_{up}$). During each current pulse event, a high driving current is followed by a small reading current to drive the magnet to the metastable state and detect the relaxed magnetization state.

However, the energy landscape would be modified by an magnetic field (***H***) through the Zeeman energy splitting of the two stable ground states, which is expressed as $-\mu_0 V M_s(\boldsymbol{M}\bullet\boldsymbol{H})$, where $\mu_0$ is the permeability of free space, $V$ is the volume of an individual magnetic entity, $M_s$ is the saturation magnetization, and ***M*** is the unit vector of the magnetization (*32*). Considering an external field ***H*** along the +$z$-direction, +$H_z$, the Zeeman energy is $E_z = -\mu_0 V M_s H_z$, and thus, the energies of the $M_z$ = +1 and −1 states shift by +$E_z$ and −$E_z$, respectively, as shown in Fig. 1C. Unless otherwise stated, a positive/negative sign represents the up/down direction, respectively, for both $H_z$ and $M_z$. Therefore, upon removal of the driving current, the magnetization relaxes to $M_z$ = +1 with a probability of $P_{up} \propto \exp(-E_z/k_B T)$, while the probability of orienting in the down state $P_{dn}$ is proportional to $\exp(+E_z/k_B T)$, under the +$H_z$. Since $P_{up} + P_{dn} = 1$, it is inferred that $P_{up} = P_{up} \big/ (P_{up} + P_{dn}) = e^{-E_z/k_B T} \big/ (e^{-E_z/k_B T} + e^{E_z/k_B T}) = 1 \big/ (1 + e^{2E_z/k_B T})$ . Note that, once the exponent $|2E_z/k_B T| < 1$, according to the piecewise linear approximation (*33*), $P_{up}$ can be deduced as:

$$P_{up} \approx -\frac{E_z}{2k_B T} + 0.5 = \frac{\mu_0 V M_s}{2k_B T} H_z + 0.5, \quad (1)$$

which is a linear function of $H_z$. This linear relationship between $P_{up}$ and $H_z$ enables the detection of $H_z$ by monitoring the probability of a relaxed state. Simply, according to equation (1), without $H$, the magnet would reorient itself to either 'up' or 'down' state with the equal probability of 50%. This stochastic switching feature has been confirmed and used for nanomagnetic logic and true random number generators in previous work (*22*, *23*).

Besides $H_z$, a $z$-component effective field $H_{z,eff}$ generated by the IP current with a collinear magnetic field through SOT can also determine $P_{up}$, making it feasible to detect an IP field with the proposed sensor. $H_{z,eff}$ is written as a function of the current density along the $x$- (or $y$-) direction $J_{x(y)}$ and the normalized magnetization $m_{x(y)}$ caused by its corresponding collinear magnetic field $H_{x(y)}$ (*19*, *24*):

$$H_{z,eff} = \frac{\hbar}{2e\mu_0 M_S t} \theta_{SH} J_{x(y)} m_{x(y)}, \quad (2)$$

where $\hbar$ is the reduced Planck constant, $e$ is the electron charge, $t$ is the thickness of the CoFeB layer and $\theta_{SH}$ is the spin Hall angle of the Ta layer. The IP field $H_x$ ($H_y$) is utilized to orient the magnetization to achieve an IP component $m_{x(y)}$ and then generate $H_{z,eff}$, as depicted in Fig. 1D (Fig. 1E). Thus, for IP field detection, based on equation (2), two conclusions can be drawn. First, for a given amplitude of $J_x$ ($J_y$), $H_{z,eff}$ depends on $H_x$ ($H_y$), whereas the direction of $H_{z,eff}$ is determined by the current polarity. Therefore, $P_{up}$ would vary with $H_x$ ($H_y$) under a given collinear IP current through the generated $H_{z,eff}$, and the variation tendencies are opposite for different current polarities. Second, $H_x$ (or $H_y$) cannot produce $H_{z,eff}$ if there is solely the transverse current $J_y$ (or $J_x$). This indicates that $P_{up}$ is invariable with the IP field under a transverse current. On the other hand, for OOP field sensing, an IP current can only excite the magnet into metastable state (high energy level) without changing the Zeeman energy, and thus $P_{up/dn}$ would vary linearly with $H_z$ under the applied IP current, where the variation tendency is independent of current polarity.

We carried out the proof-of-principle experiments with a nanomagnet comprised of Ta (10 nm)/CoFeB (1 nm)/MgO (2 nm)/Ta (2 nm) (from the bottom) heterostructure, to describe the feasibility of vector magnetic field sensing based on probabilistic nanomagnet and evaluate the performance (detectable field and resolution limit) of such a sensor. The detailed magnetic and transport properties of the device are shown in Supplementary Section 2. The CoFeB with PMA can be driven to $M_z = 0$ state if an IP current flowing through a resistive bar made of tantalum that can exert strong SOT. The measurement sequence is illustrated in Fig. 1F. Here, the magnitude and the duration of the driving current are 12 MA/cm$^2$ and 0.2s, respectively. After relaxation, a reading current of 0.5 MA/cm$^2$ is used to measure the $R_{AHE}$ and thus detect the magnetization state. The state probability is collected by repeating this measurement. For example, if we count the number of up states as $N_{up}$ in $N$ pulse event cycles, then $P_{up}$ is equal to $N_{up}/N$. Therefore, our method can detect not only DC fields but also AC fields. The maximum frequency of a detectable AC field is expected to be inversely proportional to the total measurement time and could reach over 1 kHz (Supplementary Section 3).

**One-dimensional magnetic field sensing**

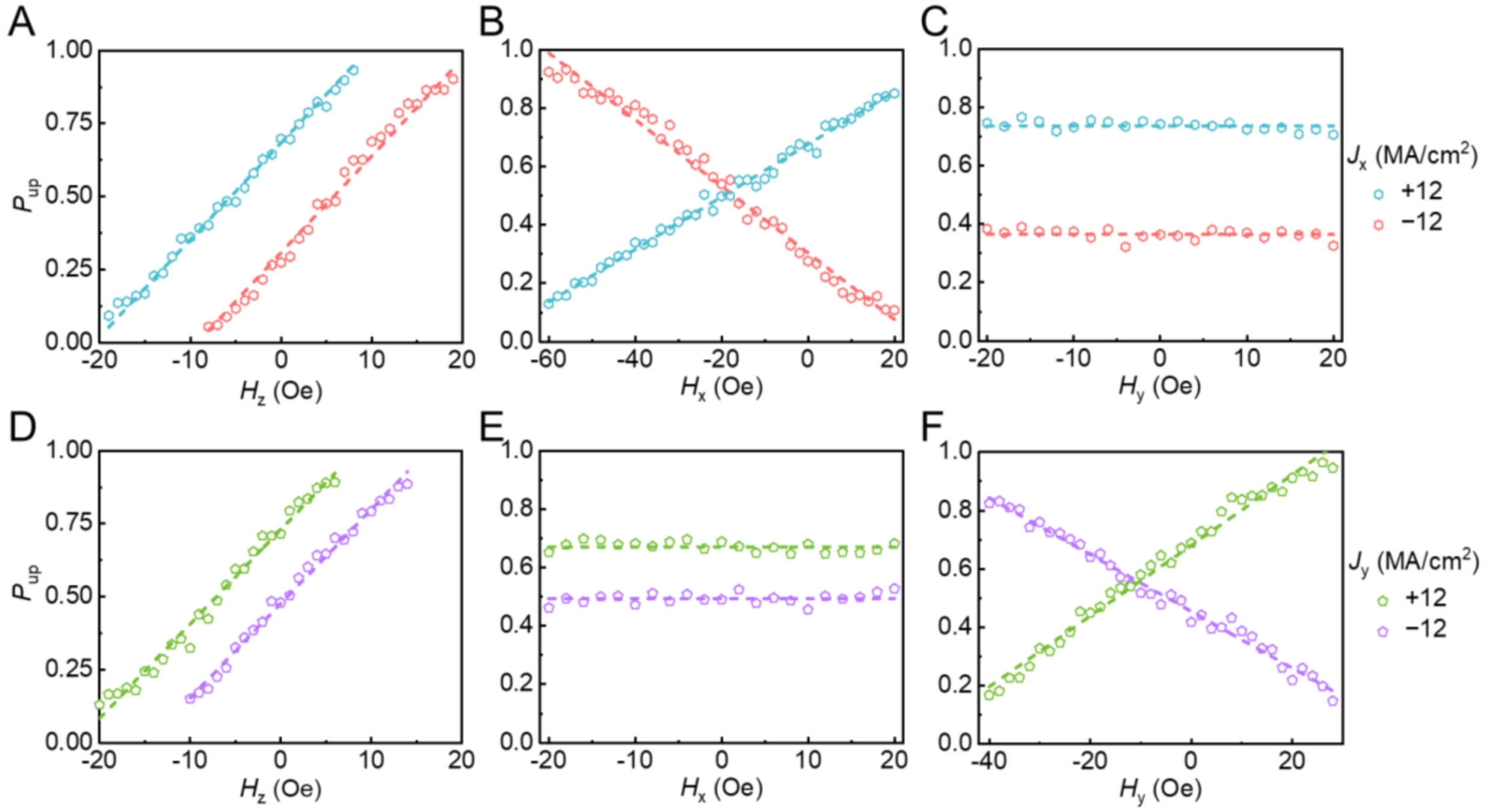

**Fig. 2. One-dimensional magnetic field sensing.** (**A** to **C**) Probability of the settled state being the up state, $P_{up}$, as a function of the magnetic field along the *z*-direction (A), *x*-direction (B), and *y*-direction (C) when a driving current pulse of 12 MA/cm$^2$/200 ms is applied along the ±*x*-direction. (**D** to **F**) $P_{up}$ as a function of magnetic field along the *z*-direction (D), *x*-direction (E), and *y*-direction (F) when the current is applied along the ±*y*-direction. The dashed lines are the linear fits to the experimental data. Each displayed measurement data point is a statistical result from 500 pulse events.

For experimental implementation of vector magnetic field sensing, we start from the detection of a one-dimensional magnetic field. Here, we apply a driving current pulse 500 times to count the up-state probability ($P_{up}$) at a given magnetic field. When the current pulse is applied along the *x*-direction ($J_x$), whether positive or negative, as discussed above, $P_{up}$ linearly increases with increasing $H_z$ from −19 to 8 Oe (or from −8 to 19 Oe), and the two cases have almost the same slope for the $P_{up}$-$H_z$ curves (Fig. 2A). In contrast, $P_{up}$ shows the opposite variation tendencies with $H_x$ for different polarities of $J_x$ (Fig. 2B). Under $J_x$ = −12 MA/cm$^2$, $P_{up}$ linearly decreases with $H_x$ within the range of −60 to +20 Oe, while under $J_x$ = +12 MA/cm$^2$, $P_{up}$ linearly increases with $H_x$. Moreover, $P_{up}$ remains constant with $H_y$ varying from −20 to 20 Oe under $J_x$, regardless of the polarity, as no $H_{z,eff}$ is induced by $H_y$, which cannot tilt the magnetization towards the *x*-direction (Fig. 2C). It is worth noting that settled up/down probabilities is unequal at $H$ = 0 Oe, and thereby existing an asymmetric linear range for $P_{up}$ vs. $H_{(x,\ y,\ \text{or}\ z)}$, which may be associated with the presence of a bias field in the above unoptimized experimental devices (Supplementary Section 4). Such nonequality and asymmetry could be improved or corrected in real applications by optimizing film deposition and device fabrication, or introducing a magnetic field to balance the bias field, or calibrating the sensor system. Overall, $P_{up}$ is found to be sensitive to $H_x$ and $H_z$, but insensitive to $H_y$ when $J_x$ is applied.

Similarly, $P_{up}$ is sensitive to $H_y$ and $H_z$, but insensitive to $H_x$ when $J_y$ is applied (Figs. 2D-F). These experimental results not only show the operation principle of one-dimensional magnetic field sensing, but also further verify the theoretical expectations described in the previous section that there are different relations between the variation tendency of the relaxed state probability and the current polarity. Next, we will investigate whether these differences enable distinction of the contributions of the IP and OOP fields. In other words, we aim to determine whether three components ($H_x$, $H_y$ and $H_z$) of the magnetic field can be detected by characterizing the relaxed state probability variation, for reconstructing vector magnetic field with one single nanomagnet.

## Three-dimensional magnetic field sensing

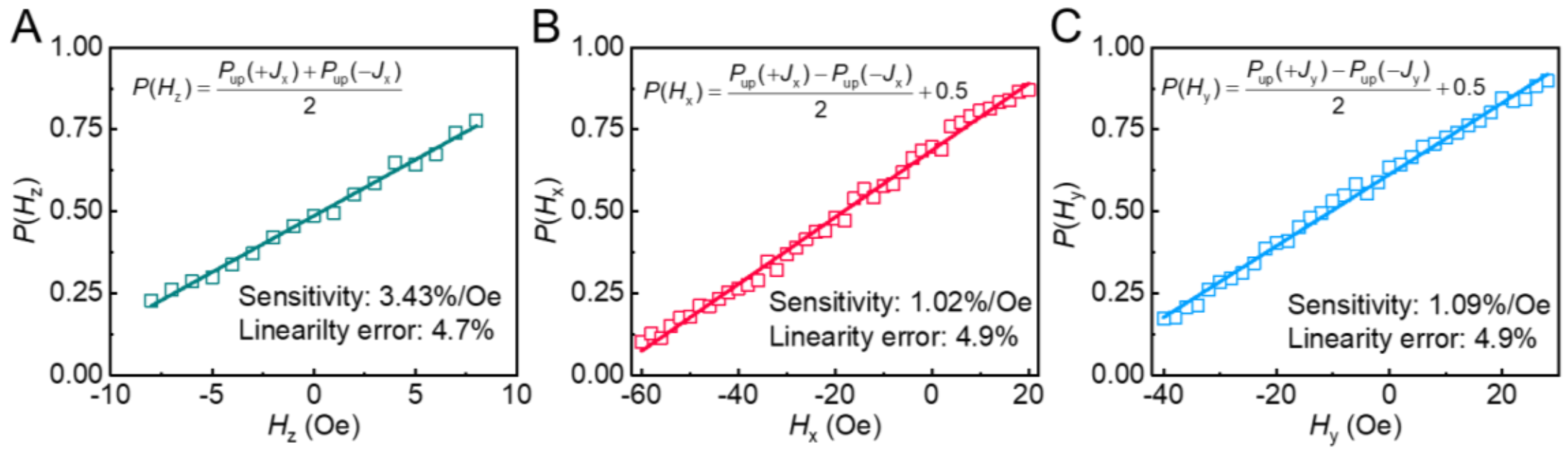

**Fig. 3. Net probability contributed by $H_z$, $H_x$ and $H_y$.** (**A**) Net probability $P(H_z)$ as a function of $H_z$, where the $P_{up}(J_x = +12\ \mathrm{MA/cm^2})$ and $P_{up}(J_x = -12\ \mathrm{MA/cm^2})$ values under different $H_z$ ranging from −8 to 8 Oe are processed with an addition operation to obtain $P(H_z)$. (**B**) Net probability $P(H_x)$ as a function of $H_x$ from −60 to 20 Oe, as obtained by a subtraction operation to eliminate the contribution of $H_z$. (**C**) The net probability component $P(H_y)$ as a function of $H_y$ in the range of −40 to +28 Oe, as obtained by a subtraction operation using the two values of $P_{up}(J_y = +12\ \mathrm{MA/cm^2})$ and $P_{up}(J_y = -12\ \mathrm{MA/cm^2})$. The solid lines are best fit to the corresponding experimental data points.

If a 3D magnetic field $\boldsymbol{H} = (H_x, H_y, H_z)$ is applied to the device, then two $P_{up}$ values under positive and negative $J_x$ can be measured: $P_{up}(+J_x)$ and $P_{up}(-J_x)$. According to the different current-direction-dependent tendencies of the $P_{up}$–$H_z$ curves shown in Fig. 2, the two collected $P_{up}$ values can be processed with an addition operation to eliminate the $H_x$ contribution. Therefore, the net probability contributed only by the $H_z$ component is shown in Fig. 3A and is expressed as:

$$P(H_z)=\frac{P_{up}(+J_x)+P_{up}(-J_x)}{2}. \tag{3}$$

By performing a subtraction operation to eliminate the $H_z$ contribution, we can obtain the net probability contributed only by $H_x$ (Fig. 3B):

$$P(H_x)=\frac{P_{up}(+J_x)-P_{up}(-J_x)}{2}+0.5. \tag{4}$$

Here, we add a compensation value of 0.5 to the subtraction result to ensure that $P_{up}(H_x)$ is in the range from 0 to 1.

Likewise, if the driving current is applied along the *y*-direction, then the net probability contributed only by $H_y$ can be evaluated as (Fig. 3C):

$$P(H_y)=\frac{P_{up}(+J_y)-P_{up}(-J_y)}{2}+0.5. \tag{5}$$

Note that $P(H_z)$ can also be calculated as:

$$P(H_z)=\frac{P_{up}(+J_y)+P_{up}(-J_y)}{2}, \tag{6}$$

and the $P(H_z)$ values calculated from equations (3) and (6) are expected to be the same.

Therefore, the relationships between the net probability ($P(H_x)$, $P(H_y)$, and $P(H_z)$) and the corresponding magnetic field components ($H_x$, $H_y$, and $H_z$) can be identified from 1D measurements, which is similar as our previous work (*19*). Once the net probability components are obtained using equations from (2) to (5), the magnitude of the corresponding magnetic field components can be read out according to Figs. 3A-C, and thus implementing 3D magnetic field sensing. The linear range of the magnetic field sensor is approximately −8 to +8 Oe for $H_z$, −60 to +20 Oe for $H_x$, and −40 to +28 Oe for $H_y$, as shown in Fig. 3. General speaking, the maximum detectable field for $H_z$ is related to the saturated field of the nanomagnet with the assistance by SOT. In contrast, the maximum detectable IP field is expected to be determined by the field required to induce deterministic switching, corresponding to 100% one-state probability, under the driving current through SOT.

**Performance of the sensor**

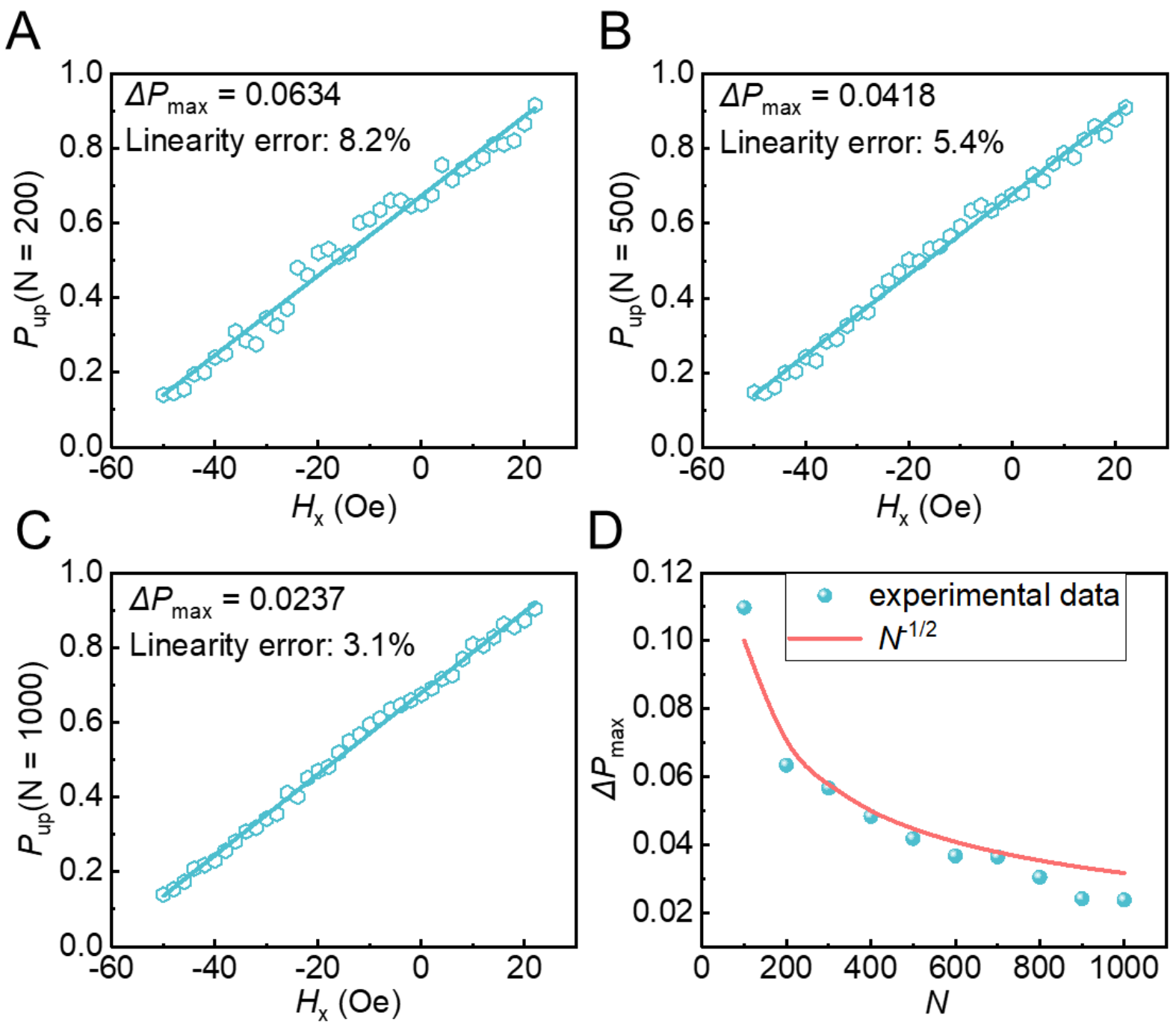

**Fig. 4. Effect of the number of pulse events on the linearity of $P_{up}$-$H_x$ curves under $J_x$ = +12 MA/cm$^2$.** (**A** to **C**) $P_{up}$-$H_x$ curves with pulse event numbers of $N$ = 200 (A), 500 (B), and 1000 (C). The cyan lines are the best fit to the experimental data points. (**D**) Linearity (cyan circles) as a function of pulse event number along with the $1/\sqrt{N}$ curve (soft red line).

Next, based on the results shown in Fig. 3, we evaluate the performances of the demonstrated nanoscale 3D magnetic field sensor, such as sensitivity, linearity, field resolution, and spatial resolution. Sensitivity is the response of the device to the field change and is given by the relation S = ΔP/ΔH. Here, the sensitivities are calculated to be 1.02%/Oe, 1.09%/Oe and 3.43%/Oe for $H_x$, $H_y$, and $H_z$, respectively. Other specification of the sensor, to measure the maximum deviation of the sensor output from the fitted straight line, is the linearity error ($\Delta P_{max}$). Normally, it is expressed as a percentage of the full-scale value: $\Delta P_{max}/P_{FS} \times 100\%$, where $P_{FS}$ is the full-scale output within the linear range (*19*). According to the results shown in Fig. 3, the linearity errors of our sensor within the linear range is 4.9%, 4.9% and 4.7% for $H_x$, $H_y$ and $H_z$, respectively. Here, $\Delta P_{max}$, along with the sensitivity, is also used to evaluate the minimum detectable field ($\delta H_{min} = \Delta P_{max}/S$) for $H_x$, $H_y$, and $H_z$, given by 3.9, 3.4, and 0.7 Oe, respectively. Furthermore, the linearity error can be further improved by increasing the number of pulse events. Figs. 4 show $P_{up}$ as a function of $H_x$ under different numbers of pulse events ($N$ = 200, 500 and 1000) with $J_x$ = +12

MA/cm$^2$. $\Delta P_{max}$ and the linearity error decrease with increasing number of pulse events and reach 2.37% and 3.1%, respectively, when $N$ = 1000 (Fig. 4C).

Generally, for all trials, the distribution of the actual $P_{up}$ after $N$ pulsing events, at a given field with expected up-state probability $P_{up}^{E}$, would obey binomial distribution (BD) with the standard deviation $\sigma = \sqrt{NP_{up}^{E}(1-P_{up}^{E})}/N$ (*34*). It indicates that $P_{up}$ is extremely likely (~95.4%) to fall into $P_{up}^{E} \pm 2\sigma$, and hence the utmost of sensor linearity error can be roughly estimated as 2σ, which reaches maximum $1/\sqrt{N}$ when $P_{up}^{E} = 0.5$ for a fixed $N$, but decreases with the increase of $N$. Fig. 4D plots $\Delta P_{max}$ as a function of $N$ and the fitting with the $1/\sqrt{N}$ curve (soft red line). It is seen that $\Delta P_{max}$ is closed to $1/\sqrt{N}$, though the former is less than the latter when $N$ is greater than 800. Consequently, the minimum detectable magnetic field $\delta H_{min}$ would be $\Delta P_{max}/S = 1/S\sqrt{N}$. Accordingly, $\delta H_{min}$ is expected to be as low as 1 μT, which is the magnitude of the magnetic field produced by the spin of a single electron at a distance of 10 nm, when $N = 7.5 \times 10^7$, $5.6 \times 10^7$, and $0.3 \times 10^7$ for the *x*-, *y*-, and *z*-directions, respectively. Especially, the $N$ for the *z*-direction is on the same order as the number of pulse events used in NV measurements (*13*). Furthermore, inspired by the NV based sensors (*12, 13*), we deduce the field resolution $\eta$ as the product of $\delta H_{min}$ and the square root of measurement time, *i.e.* $\eta = \mu_0 \delta H_{min}\sqrt{NT_0}$, where $T_0$ is the time of each current cycle. The long cycle period ($T_0$ = 0.3 s) used in our conceptual experiments leads to a large η (4.8, 4.2, and 0.8 $\mathrm{mT}/\sqrt{\mathrm{Hz}}$ for the *x*-, *y*-, and *z*-directions, respectively, at $N$ = 500). $T_0$ could reach 30 ns for real applications (*35, 36*), and consequently, the resolution would theoretically be improved to 1.5, 1.3, and 0.3 $\mu\mathrm{T}/\sqrt{\mathrm{Hz}}$.

Additionally, we expect that the spatial resolution can be developed by further scaling down the nanomagnet size. Although a smaller magnetic body volume $V$ will decrease the energy barrier, it has been demonstrated to tune the switching probability of low-barrier nanomagnets, such as $\Delta = K_u V/k_B T \approx 9$, where $K_u$ is magnetic anisotropy constant, by both OOP magnetic fields and IP fields (*27, 37*). In other words, the proposed sensor could keep working even with such low-barrier magnets for detecting both OOP fields and IP fields, giving us an opportunity to promote the spatial resolution of the sensor to less than 10 nm (see Supplementary Section 2 more details). Also, scaling down the size of the nanomagnet would reduce the driving current and thus power consumption. It indicates this sensing technique is to allow the sensor to be scalable, which is crucial for nanoscale magnetic detection and imaging with high spatial resolution. For future real application, the proposed sensor with Hall bar structure could be developed into a magnetic tunnel junction (MTJ) structure that would have more advantages, for example, a larger readable resistance ratio due to the TMR effect (*38*). In future, to implement simultaneous magnetic imaging over a wide field of view such as μm or mm order, an MTJ array could be designed with a shared common driving path to enhance the efficiency in imaging the spatial magnetic field (*39*). Although individual MTJs are separated from each other, it would be possible to obtain continuous spatial field imaging with two layers of interleaved sensors, by making use of the advanced technique of 3D stacking of MTJs (Supplementary Section 5).

The field resolution and spatial resolution of our sensor are comparable to the quantum vector magnetometers such as single NV based sensors (*17*, *18*), if the promoted read/write speed and size of our device are utilized (see Supplementary Section 6 for more details). More importantly, our technique has advantages such as one single device structures with all-electric operations. These properties will allow us to develop a simple and compact sensor system for detecting vector magnetic fields at nanoscale dimension.

## Discussion and outlook

It is expected that the driving and reading of our proposed sensor device could be carried out in the same way as already used in SOT-magnetoresistive random-access memory technology (*40*). Moreover, nanomagnet chips can be integrated with not only electronics but also microfluidics for biomedical engineering applications (*41*). Finally, with the same nanomagnet, the demonstrated possibility of sensing nanoscale magnetic fields or electrical current in combination with the observed memory and stochastic or probabilistic computing capabilities (*21-23*) opens a bright new world of nanospintronics for integrated sensing, memory and computing functions.

## Acknowledgments:

**Funding:** This work was supported by the National Natural Science Foundation of China (NSFC Grant Nos. 62074063, 61821003, 61904060, 61904051, 51671098, and 61674062). This work was supported by National Key Research and Development Program of China (No. 2020AAA0109005), the Interdisciplinary program of Wuhan National High Magnetic Field Center (No. WHMFC202119), Huazhong University of Science and Technology, and Fund from Shenzhen Virtual University Park (No. 2021Szvup091).

**Author contributions:**

Conceptualization: L.Y.

Methodology: L.Y., S.Z., S.L., X.Y., X.Z.

Investigation: L.Y., S.L., S.Z., Y.X., R.L.

Visualization: S.Z., S.L., Z.C., M.S.

Funding acquisition: L.Y.

Project administration: L.Y.

Supervision: L.Y.

Writing – original draft: L.Y., S.Z., S.L.

Writing – review & editing: L.Y., S.Z., S.L., Z.G.

**Competing interests:** Authors declare that they have no competing interests.

**Data and materials availability:** All data are available in the main text or the supplementary materials and are available on request.

## Supplementary Materials

Materials and Methods

Supplementary Text

Figs. S1 to S6

Tables S1 to S2

References (*42–57*)

# Supplementary Materials for

## Nanoscale three-dimensional magnetic sensing with a probabilistic nanomagnet driven by spin-orbit torque

Shuai Zhang[1]†, Shihao Li[1]†, Zhe Guo[1], Yan Xu[1], Ruofan Li[1], Zhenjiang Chen[1],

Min Song[2], Xiaofei Yang[1], Liang Li[3], Jeongmin Hong[1], Xuecheng Zou[1], Long You[1,3]*

*Corresponding author. Email: lyou@hust.edu.cn

†These authors contributed equally to this work.

**This PDF file includes:**

**Materials and Methods**

**Sample preparation** Magnetron sputtering was used to deposit a film structure of Ta(10 nm)/CoFeB(1 nm)/MgO(2 nm)/Ta(2 nm) on a thermally oxidized Si substrate at room temperature. The film was then processed into the Hall-bar structure by electron beam lithography (EBL) and argon-ion milling (AIM). The Hall bars contained the entire thin film stack, with the region outside the Hall bars etched down to the insulating Si substrate. We used EBL and AIM to define the current channel and the detection channel. The widths of the two channels are designed to be 1 μm. A 10 nm thick hard mask (Ti) with sizes of 200 × 200 nm$^2$ were grown at the center of the Hall bars by EBL and deposited by electron beam evaporation. AIM was also used to etch the stack outside the dot's region down to the bottom Ta layer. The dots therefore comprised Ta (10 nm)/CoFeB (1 nm)/MgO (2 nm)/Ta (2 nm), and the regions of the Hall bar outside the dots were etched down to the bottom Ta layer.

**Electrical measurements** For the anomalous Hall resistance measurements, we used a d.c. current source (Keithley model 6221) to apply currents and a nanovoltmeter (Keithley model 2182A) to measure the Hall voltage. A high current of 1.2 mA (12 MA cm$^{-2)}$ with a duration of 0.2 s was used excite the nanomagnet mentioned in the main text. A constant low current of 50 μA (0.5 MA cm$^{-2}$) was applied to read out the AHE resistance. The external magnetic field was generated by a Helmholtz coil driven by a power supply (Model EM5 & Model P7050, East Changing Technologies, Inc. Beijing).

**Supplementary Text 1: Deduction of the probability as a function of fields through Arrhenius-Néel law**

We consider a binary nanodevice comprising of heavy metal/ferromagnet (HM/FM) bilayer with perpendicular magnetic anisotropy (PMA). As shown in Fig. S1A, the up and down states ($M_z$ = ±1) of the binary nanomagnet are separated by an energy barrier $E_b$ such that the stored information can be retained for a time $\tau = \tau_0 e^{\frac{E_b}{k_B T}}$, following Arrhenius' law, where $\tau_0$ is the attempt time ($\tau_0 \approx 1$ ns) (*42*), $k_B$ the Boltzmann constant, and $T$ the temperature. In the following text, we will deduce the probability of being in a stable state as a function of out-of-plane field by Arrhenius law and Neel-Brown model (*42, 43*):

**① deduction using Arrhenius law**

The energy profile would be modified by an out-of-plane (OOP) magnetic field $H_z$ due to the Zeeman energy $E_z = -\mu_0 V M_s H_z$, where $\mu_0$ is the permeability of free space, $M_s$ the saturation magnetization and $V$ the volume of individual magnetic entity (*44, 45*). Then, the energies of the $M_z$ = +1 and −1 state shift by $E_z$ and $-E_z$, respectively. In other words, the energy barrier from down state to up state becomes $E_b$+$E_z$, while the energy barrier from up state to down state becomes $E_b$−$E_{zz}$, as illustrated in Fig. S1B. As a result, the dwell times of being in down and up states become different: $\tau_{dn} = \tau_0 e^{\frac{E_b+E_{zz}}{k_B T}}$ and $\tau_{up} = \tau_0 e^{\frac{E_b-E_{zz}}{k_B T}}$. The probability of being in up state $P_{up}$ can be described as (*46*)

$$P_{up} = \frac{\tau_{up}}{\tau_{dn}+\tau_{up}} = \frac{1}{1+e^{\frac{2E_z}{k_B T}}}\,, \tag{S1}$$

which is the same form as the equation deduced in the main text.

**② deduction by Neel-Brown model**

According to Neel-Brown model (*42, 43*), the dwell times are given by $\tau_{dn} = \tau_0 e^{\frac{E_b}{k_B T}(1-\frac{H_z}{H_k})^2}$, and $\tau_{up} = \tau_0 e^{\frac{E_b}{k_B T}(1+\frac{H_z}{H_k})^2}$,where $E_b = \mu_0 V M_s H_k/2$, and $H_k$ is the effective magnetic anisotropy field. Then, $P_{up}$ is expressed as $P_{up} = \frac{\tau_{up}}{\tau_{dn}+\tau_{up}} = \frac{1}{1+e^{-\frac{2\mu_0 V M_s H_z}{k_B T}}} = \frac{1}{1+e^{\frac{2E_z}{k_B T}}}$, the same form as the equation (S1).

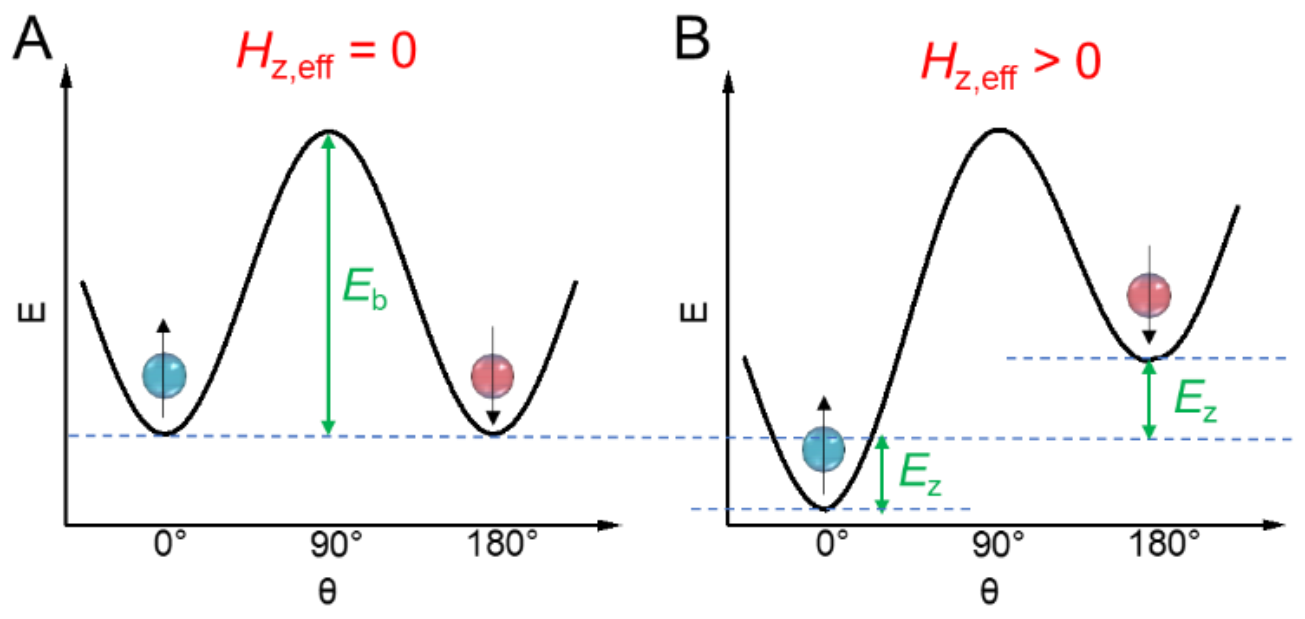

**Fig. S1. Energy profile of the magnetization under (A) zero field or (B) an out-of-plane $H_{z,eff}$.**

**Supplementary Text 2: Magnetic and transport properties of the sensor device.**

Magnetic transport characterizations of the nanodevice are shown here. Fig. S2A shows the *M*-*H* loop of the multilayer film along out-of-plane direction under room temperature by Physical Property Measurement System (PPMS). The coercivity field and the saturation magnetization of the film are around 10 Oe and 750 emu/cc, respectively. The film was then processed to a nano dot with a size of 200 nm × 200 nm sitting on a Hall bar. The anomalous Hall effect (AHE) loop under a small reading current of 0.5 MA/cm$^2$ (Fig. S2B) shows a sharp switching of the resistance, indicating a good PMA of the device with a coercivity field of ~60 Oe and an anisotropy field, $H_k$, of 3000 Oe (Fig. S2C). Accordingly, the anisotropy energy density is estimated as $K_u = \mu_0 M_s^2/2 + \mu_0 M_s H_k/2 = 4.66 \times 10^5$ J/m$^3$. If the magnet body volume $V$ decreases to 9 nm × 9 nm × 1 nm, the thermal stability factor $\Delta = K_u V/k_B T \approx 9$, with which the nanomagnet can keep working to sense magnetic fields. Fig. S2D shows the $R_{AHE}$-$J_x$ switching curve with external $H_x$ = 200 Oe, to verify the switching is mainly induced by SOT effect.

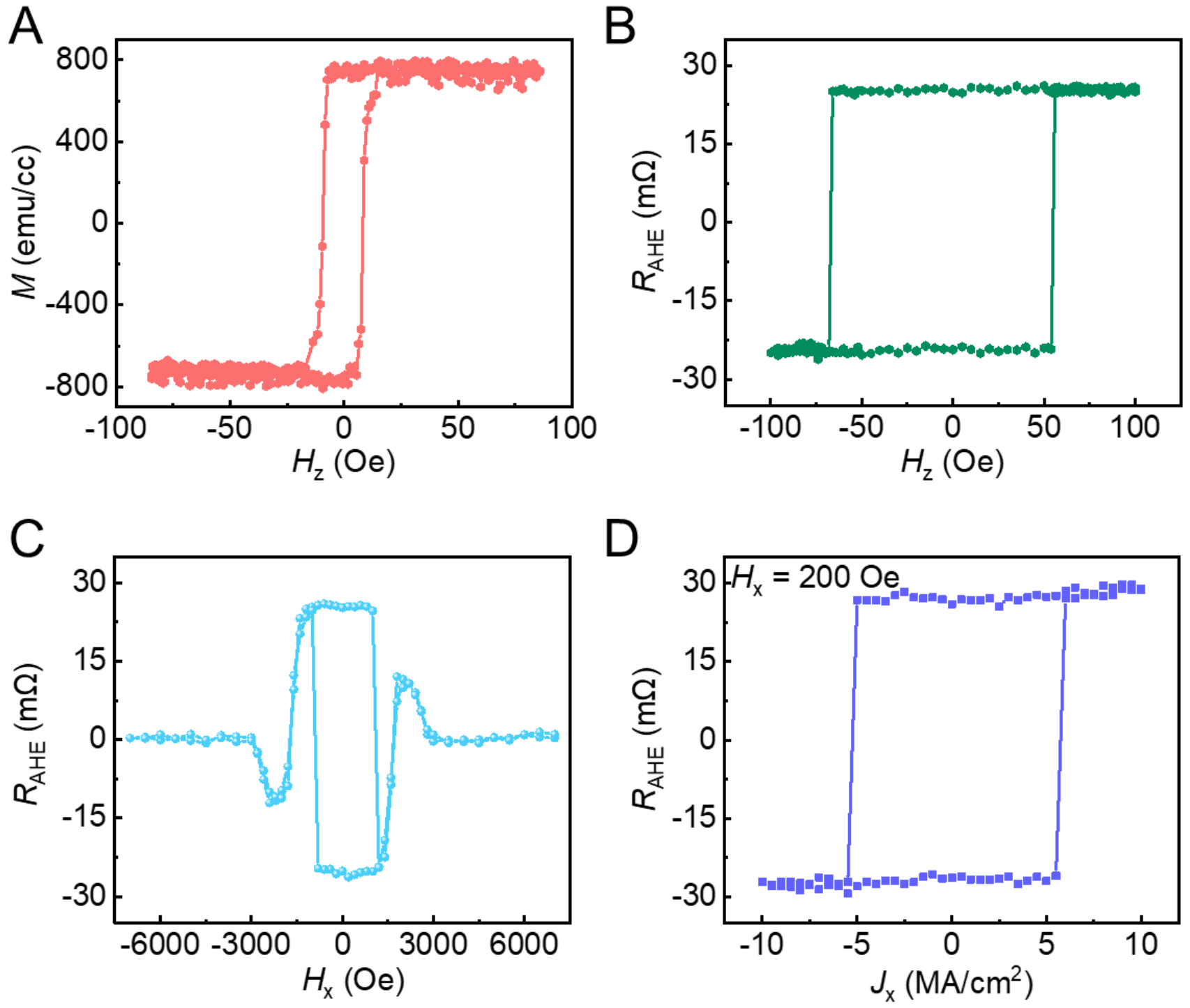

**Fig. S2. Magnetic and transport properties of the sensor device.** (**A**) *M*-*H* loop of the multilayer film along OOP direction. (**B**) Deterministic switching of the magnetization induced by OOP fields $H_z$. The reading current is 0.5 MA/cm$^2$. (**C**) AHE resistance as a function of IP fields $H_x$. (**D**) Deterministic switching of the magnetization induced by currents with the assistance of a small IP field of 200 Oe due to SOT.

## Supplementary Text 3: Measurement method of detecting an AC vector field

Here, we discuss the possibility for detecting of AC vector fields by the sensor. Considering a cosine magnetic field applied on the device, we can measure the AC field using the piecewise approximation method. As illustrated in Fig. S3, an AC field can be approximated as several different DC fields with a duration of $T$, which should be no shorter than the measurement time of the detection of a DC vector field, *i.e.* ~$4NT_0$, where N is the numbers of current pulse events and $T_0$ the time of each current pulse cycle. On the other hand, when a cosine wave is staircase approximated, the number of segments must be no less than 12 to guarantee the fidelity of the detected AC field (*47*). Note that the fidelity would be higher as increasing the segment. Therefore, a frequency below $f_H = 1/(12\times4NT_0)$ can be detected with an acceptable accuracy, and the fidelity is higher at the lower frequency. Assuming $N = 500$, $T_0 = 30$ ns, then $f_H \approx 1.39$ kHz, which means that the proposed sensor can detect a vector magnetic with a frequency ranging from DC to 1.39 kHz.

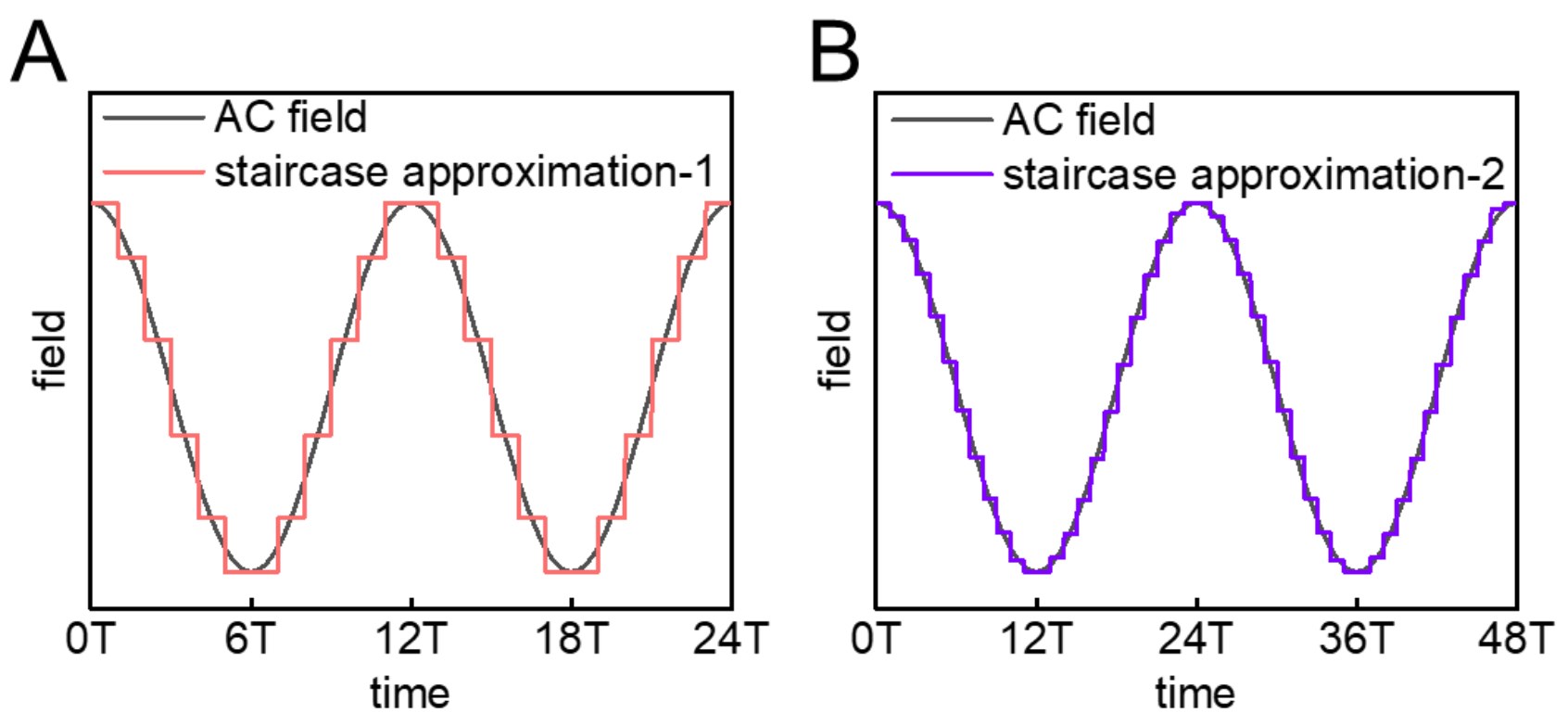

**Fig. S3. Detecting AC magnetic fields by staircase approximation.** (**A**) A cosine field (black line) is cut into 12 segments (orange line) during each cycle. (**B**) A cosine field (black line) is cut into 24 segments (purple line) during each cycle. Each segment should be no shorter than the measurement time of the detection of a DC vector field, *i.e.* $4NT_0$.

## Supplementary Text 4: Calculated *P–H* curves

Fig. S4 shows the net probability $P(H_z)$–$H_z$ curve when the current is applied along *y*-direction, as obtained by an addition operation on the two $P_{up}$–$H_z$ curves at $J_y$ = +12 MA/cm$^2$ and $J_y$ = −12 MA/cm$^2$ (Fig. 2D of the main text). $P(H_z)$ increases linearly with increasing $H_z$ with a sensitivity of 3.5%/Oe and a linearity error of 4.7%, which is consistent with the $P(H_z)$–$H_z$ curve when current is along *x* direction (Fig.3A of the main text).

For the two curves shown in Fig.2A of the main text, besides an addition operation, a subtraction operation can be also performed to eliminate the $H_z$ contribution, and then $P(H_x)$–$H_z$ is obtained, as shown in Fig. S5A. The $P(H_x)$ values does remain a constant as expected, although the constant deviates a little from the ideal value of 0.5. In the ideal situation, the probability $P_{up}$ would be 0.5 at zero field, as reported by our previous work (*48*). However, the field corresponding to $P_{up}$ = 0.5 shifts from zero field (Fig.2 of the main text). These deviations may be caused by a bias field ($\boldsymbol{H}_b$ = ($H_{b,x}$, $H_{b,y}$, $H_{b,z}$)) within the device. The bias field also results in the asymmetries of linear ranges in Fig. 3 of the main text. Taking Fig. 3A as an example, an external field along *z* direction of about 0.4 Oe can balance the $H_{b,z}$ and obtain a probability of 0.5, indicating $H_{b,z}$ ≈ –0.4 Oe. Similarly, $H_{b,x}$ ≈ 18.3 Oe and $H_{b,y}$ ≈ 10.3 Oe are evaluated according to Fig. 3B and 3C of the main text, respectively. In addition, under the existences of $H_{b,x}$, the sign of the produced $H_{z,eff}$ is opposite if the current reverses according to Equation (2) of the main text, and thus the two curves at $J_x$ = +12 MA/cm$^2$ and $J_x$ = −12 MA/cm$^2$ does not coincide with each other, both for Fig. 2A and 2C of the main text. It means that, if the bias field is eliminated by optimizing film deposition and device fabrication, the two curves under the positive and negative currents both in Fig. 2A and 2C should be entirely coincidence and the $P_{up}$ = 0.5 at zero field, while the two curves in Fig. 2B should intersect at the point of ($P_{up}$ = 0.5, $H_x$ = 0).

Figs. S5A-D depict the curves of $P(H_x)$–$H_z$, $P(H_x)$–$H_y$, $P(H_z)$–$H_x$, and $P(H_z)$–$H_y$, where $P(H_z)$ and $P(H_x)$ are calculated with the measured $P_{up}(\pm J_x)$ in Figs. 2A-C of the main text. Therefore, by performing a subtraction operation, $P(H_x)$ linearly varied with $H_x$ (Fig.3B of the main text) but is unchanged (~0.5) with varying $H_z$ (Fig. S5A) or $H_y$ (Fig. S5B); by performing an addition operation, $P(H_z)$ changes linearly with $H_z$ (Fig.3A of the main text) but remains a constant close to 0.5 with changing $H_x$ (Fig. S5C) or $H_y$ (Fig. S5D). In addition, the curves of $P(H_y)$–$H_z$, $P(H_y)$–$H_x$, $P(H_z)$–$H_x$, and $P(H_z)$–$H_y$ are depicted in Figs. S5E-H, where $P(H_y)$ and $P(H_z)$ are calculated with the measured $P_{up}(\pm J_y)$ in Figs. 2D-F of the main text. Similarly, these four curves are almost a constant about 0.5, conforming that the net probability is insensitive to the orthogonal fields.

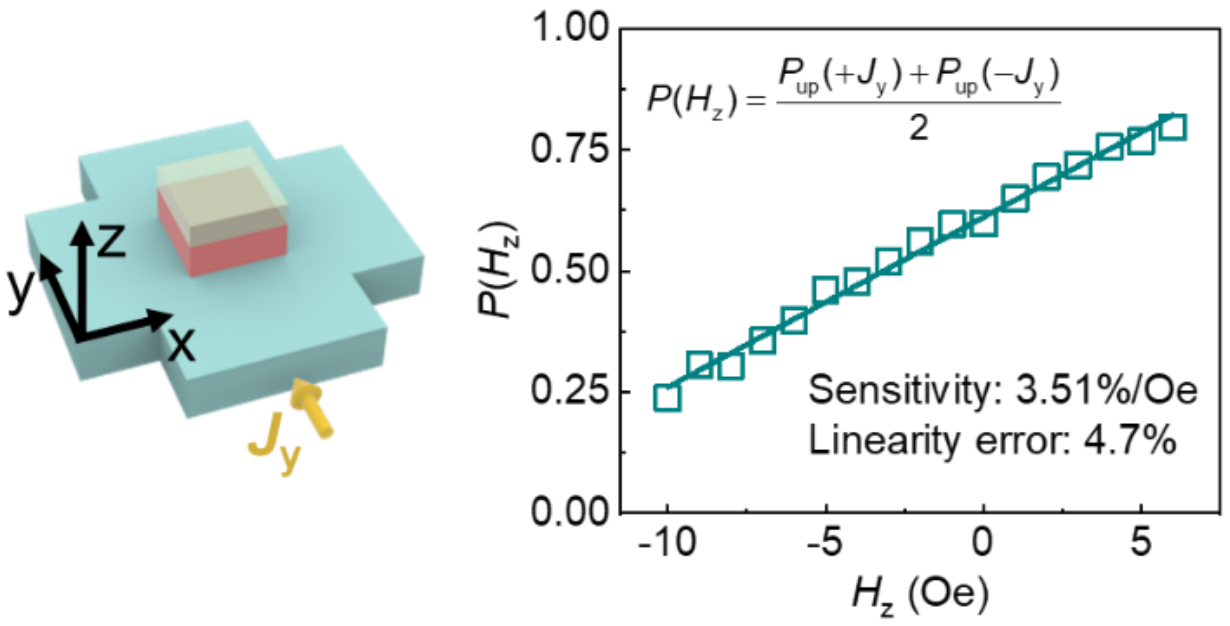

**Fig. S4. Net probability *P*($H_z$) as a function of $H_z$ when the current is along the *y*-direction.**

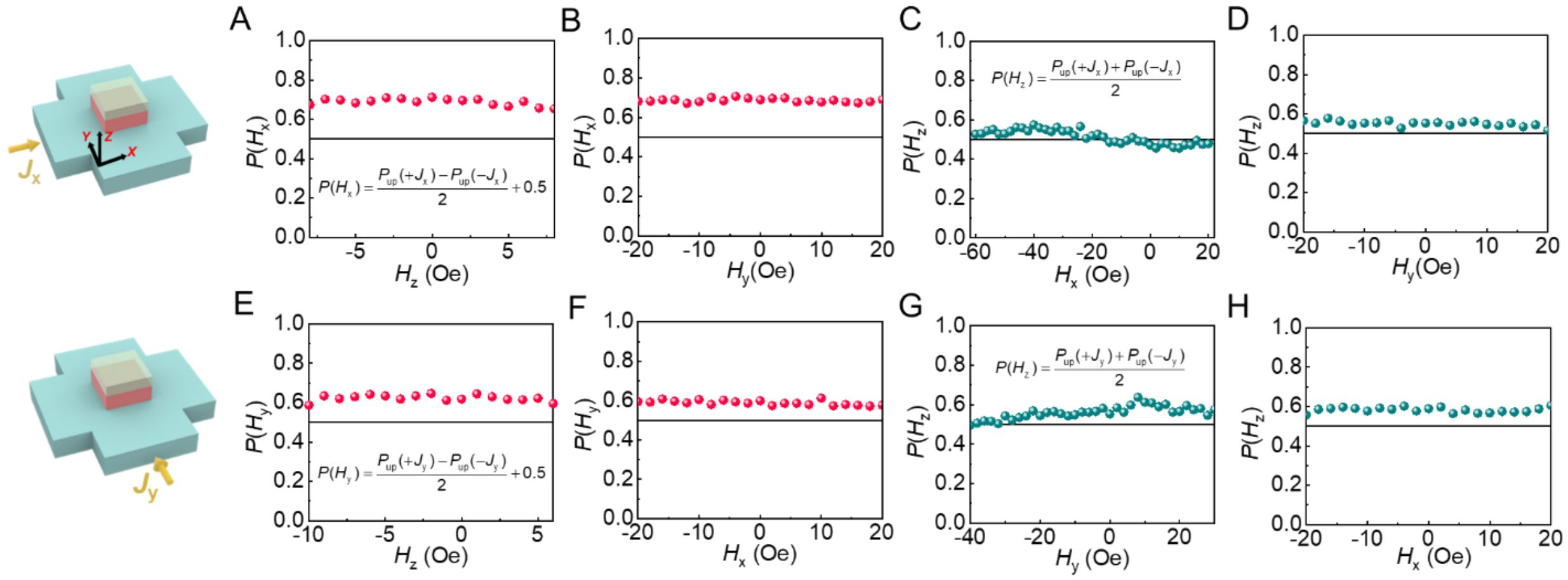

**Fig. S5. Net probability as a function of the magnetic fields along the insensitive directions.**

## Supplementary Text 5: Circuit scheme of the CMOS/sensor array

For practical applications, the proposed sensor would be developed to MTJ structure to achieve more readable resistances due to TMR effect. This single sensor can be attached to a tip of a scanning probe. However, when mapping a magnetic field distribution in space or the surface of magnetic samples, the method of scanning probe is undoubtedly inefficient. We design a MTJ sensor array located at the center of a cross bar of heavy metal layer to realize fast magnetic field mapping, as illustrated in Fig. S6A. All the MTJs share the driving path, but the resulted states, after applying one single pulse of driving current, of each MTJ are read and counted individually. To be specific, one single driving current pulse can drive all MTJs to the metastable state, and when the driving current is cut off, MTJs are read column by column activated by the WL. The total measurement time can be greatly reduced by this approach. Therefore, through sharing the driving current channel, a high-speed and high-density sensor array can be achieved.

The driving control circuits and read circuits, along with the pulse sequence of the controlled signals, are illustrated in Fig. S6B and S6C, respectively. Here are the detailed working processes of the sensor array (it is assumed that 512 cycles of driving and reading current are applied to a single device for all the four current directions: +X, −X, +Y and −Y direction.):

① The control signals consist of a clock signal for driving (CLKD) and a reading CLKR, where the falling of CLKD is followed by the rising of CLKR during each cycle. In order to change the current direction after every 512 pulses of CLKD are applied, we add a 11-bit counter, of which the input is the CLKD and the output signals of Q10-Q8 are also acting as control signals parameterized as A-C, respectively, along with a driving control circuits, as illustrated in the left panel of Fig. S6B. In addition, the reset signal of counter CLR is determined by the falling of signal C.

② According to the driving control circuits, the driving current direction is determined by the levels of both A and B, as summarized in Table S1.

③ After each driving pulse is applied, a reading current pulse is enabled by the signal CLKR. The MTJ state is detected by a comparator with a reference voltage, and then the number of high-resistance states $N_H$ is counted by a counter. After applying N = 512 cycles of pulses to the single device, $N_H$ will be transformed to Nx+ (Nx−, Ny+, or Ny−) if the driving current flows along +X (−X, +Y, or −Y) direction, which is used to be calculated to get the 3D magnetic field vector.

However, this array has many spaces between each two adjacent magnets which cause a missing signal of the detected spatial magnetic fields. We further develop the array by the 3D stacking technology, as illustrated in Fig. S6D. The HM layer is sandwiched between the two groups of stacks, ensuring that the two layers of interleaved sensors implement the simultaneous magnetic imaging of a spatial magnetic field over a wide-field of view such as mm order.

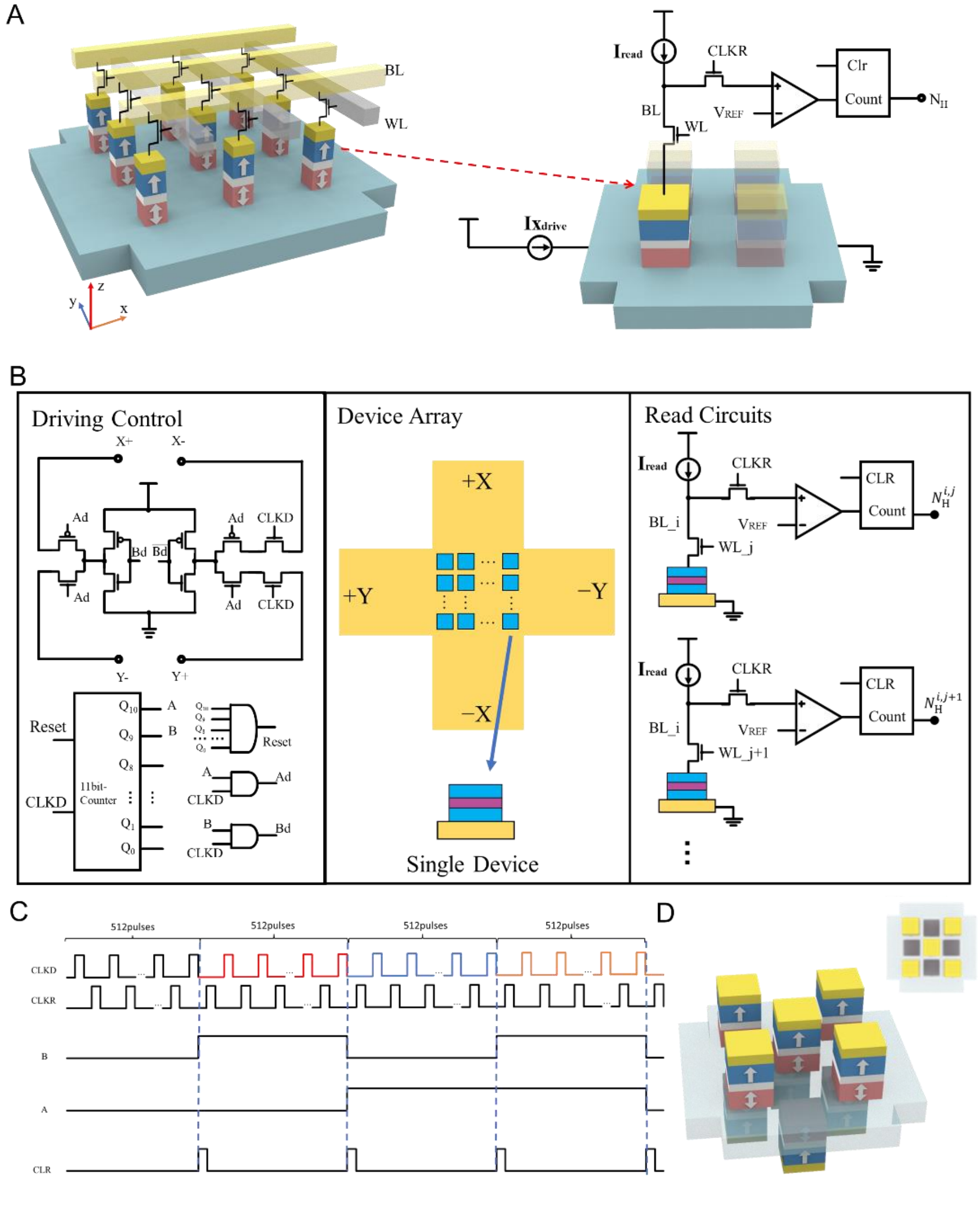

**Fig. S6. Schematic of sensor array along with the driving and reading circuits.** (**A**) Sensor array along with CMOS transistors. The magnet (red cube) adjacent to the heavy metal layer act as sensing units to detect the magnetic field and also constitute the free layer in the MTJ. The right panel shows the driving and reading circuits configuration for a single cell in the array. Each sensor is selected by a transistor with a bit line BL and a word line WL. The state of sensing magnet is detected by the resistance of the MTJ due to TMR effect. After applying N (512, for example) cycles of pulses to the single device, the number of high-resistance states $N_H$, which is used to be calculated to get the 3D magnetic field vector, is counted by a counter along with a comparator. (**B**) Schematic of the sensor array along with the detailed driving control circuits and read circuits. (**C**) Pulse sequence of the controlled signals including clock signal for driving CLKD and reading CLKR, the output signals of the 11-bit counter A,B, and the reset signal of counter CLR. It is assumed that 512 cycles of driving and reading current are applied to a single device for all the four directions: +X, −X, +Y and −Y direction. (**D**) A compact sensor configuration for magnetic field measurements with high spatial resolution, where two layers of interleaved sensors measure the magnetic field simultaneously without missing any area in the plane.

| Signal A | Signal B | Current direction |
|---|---|---|
| **0** | **0** | +X → −X |
| **0** | **1** | −X → +X |
| **1** | **0** | −Y → +Y |
| **1** | **1** | +Y → −Y |

**Table S1. driving current direction determined by signal A and B.**

**Supplementary Text 6: Performance comparison between our sensor and other types of sensors**

In Table S2, we compare key data for our sensor with state-of-the-art magnetic field sensing technologies including nanoscale graphene-based Hall sensor, scanning superconducting quantum interference device (SQUID) technology, and nitrogen/silicon vacancy (NV/SiV) based sensors. As shown in Table S2, although the 1D magnetic field sensors such as Hall sensor (row 2), scanning SQUID (row 3) and single NV based 1D sensor (row 4) show slightly better spatial resolutions (column 3), the size of our sensor (200 nm, here) is capable to be further scaled down to sub-10 nm. On the other hand, these magnetic field sensors (rows 2-4) can solely measure the magnitude of the component of magnetic field that lies along its sensitive axis.

The distribution of NV orientations (typically in NV ensembles) along multiple crystallographic axes in diamond has been explored to measure vector fields (rows 5 and 6), and can reach excellent field resolutions, but nanoscale resolution is difficult to achieve in such ensemble-based sensors. When compared with the single NV based 3D magnetic sensor technology (rows 7 and 8), the field resolution and spatial resolution of our sensor are comparable to them, if the optimized read/driving speed of $T_0$ = 30 ns and device size of 9 nm are utilized. Finally, the field resolution of the single SiV 3D sensor (row 9) is not as good as our technology in the optimized case. More importantly, when compared to the NV/SiV based techniques, the measurement method of our sensor is much simpler thanks to no requirement of magnetic field, microwave, or optical equipment.

| Technology | Measured field | Sensor head size (nm) | Field resolution | Measurement method |
|---|---|---|---|---|
| Graphene-based Hall sensor (*49*) | 1D, DC | 85 nm | 59 µT/ Hz$^{1/2}$ | Hall effect |
| Scanning SQUID (*50*) | 1D, DC | 150 nm | 76 nT/ Hz$^{1/2}$ | Josephson effect and flux quantization |
| Single NV (*51*) | 1D, DC/AC | 10-20 nm | 18nT/ $Hz^{1/2}$ (ac);<br>72nT/ $Hz^{1/2}$ (dc); | Optically detected magnetic-resonance (ODMR) |
| Ensemble of NVs (*52*) | 3D, DC | 2 mm;<br>2 mm;<br>NA; | 1nT $Hz^{1/2}$ ($\eta_x;\eta_y;\eta_x;$) | ODMR of an ensemble of NV centers along four crystallographic orientations |
| Ensemble of NVs (*53*) | 3D, AC | 4 mm;<br>4 mm;<br>0.5 mm; | 50pT/ $Hz^{1/2}$ ($\eta_x;\eta_y;\eta_x;$) | ODMR of an ensemble of NV centers along four crystallographic orientations |
| Single NV (*54*) | 3D, DC | 10-20 nm*;<br>10-20 nm*;<br>10-20 nm*; | 0.56 μT/ $Hz^{1/2}$ ($\eta_\perp$);<br>0.42 μT/$Hz^{1/2}$ ($\eta_z$); | Ramsey interferometry and quantum lock-in detection with an ancillary nuclear spin under the assistance of magnetic fields |
| Single NV (*55*) | 3D, AC | 10-20 nm*;<br>10-20 nm*;<br>10-20 nm*; | 1.1 μT/ $Hz^{1/2}$($\eta_x$);<br>0.95 μT/ $Hz^{1/2}$($\eta_z$);<br>9°/ $Hz^{1/2}$($\eta_{\theta g}$); | Rotating-frame Rabi magnetometry |
| Single SiV (*56*) | 3D, DC | NA; | 200 μT/ $Hz^{1/2}$($\eta_B$);<br>30°/ $Hz^{1/2}$($\eta_\theta$); | ODMR with reference fields |
| This work<br>(single nanomagnet) | 3D, DC/AC | 200**;<br>200**;<br>1; | 4.8 mT/ $Hz^{1/2}$($\eta_x$);<br>4.2 mT/ $Hz^{1/2}$($\eta_y$);<br>0.8 mT/ $Hz^{1/2}$($\eta_z$);<br>($T_0$ = 0.3 s)<br>1.5 μT/$Hz^{1/2}$($\eta_x$);<br>1.3 μT/$Hz^{1/2}$($\eta_y$);<br>0.3 μT/$Hz^{1/2}$($\eta_z$);<br>(assuming $T_0$=30ns) | Repeating driving and reading current pulses |

Note: NA: Not applicable

*: The exact sizes are not available from the articles, but are referred to other works (*51*, *57*).

**: The sizes of our nanomagnet as expected could be promoted to 9 nm × 9 nm (Supplementary Text 2).

**Table S2. Performance comparison between our sensor and other types of magnetic field sensors.**